\input{aipcheck}

\documentclass[
    ,final            
  ]
  {aipproc}

\layoutstyle{6x9}
\usepackage{wrapfig}

\begin{document}

\title{Forward physics with the LHCb experiment\footnote{presented at DIFFRACTION 2012, Puerto del Carmen, Lanzarote, Spain, September 10-15th, 2012.}}


\author{Dmytro Volyanskyy~(on behalf of the LHCb collaboration)}{
  address={Max-Planck-Institut f\"ur Kernphysik, PO Box 103980, 69029 Heidelberg, Germany}
}

\begin{abstract}
Due to its unique pseudorapidity coverage and the ability to perform
measurements at low transverse momenta $p_{\rm T}$, the LHCb detector
allows a unique insight into particle production in the forward
region at the LHC. Using large samples of proton-proton collision data 
accumulated at $\sqrt{s}=7$~TeV, the LHCb collaboration has performed 
a series of dedicated analyses 
providing important input to the knowledge of the parton density functions, 
underlying event activity, low Bjorken-$x$ QCD dynamics and exclusive processes. 
Some of these are briefly summarised here.
\end{abstract}

\maketitle


\section{Forward energy flow}
The energy flow $dE_{\rm total}/d\eta$ created in high-energy 
hadron-hadron collisions at large values of pseudorapidity $\eta$
is expected to be directly sensitive to the amount of parton radiation 
and multi-parton interactions~\cite{PhysRevD.36.2019}.
\begin{figure}[b!]
\centering
\includegraphics[width=0.295\textwidth]{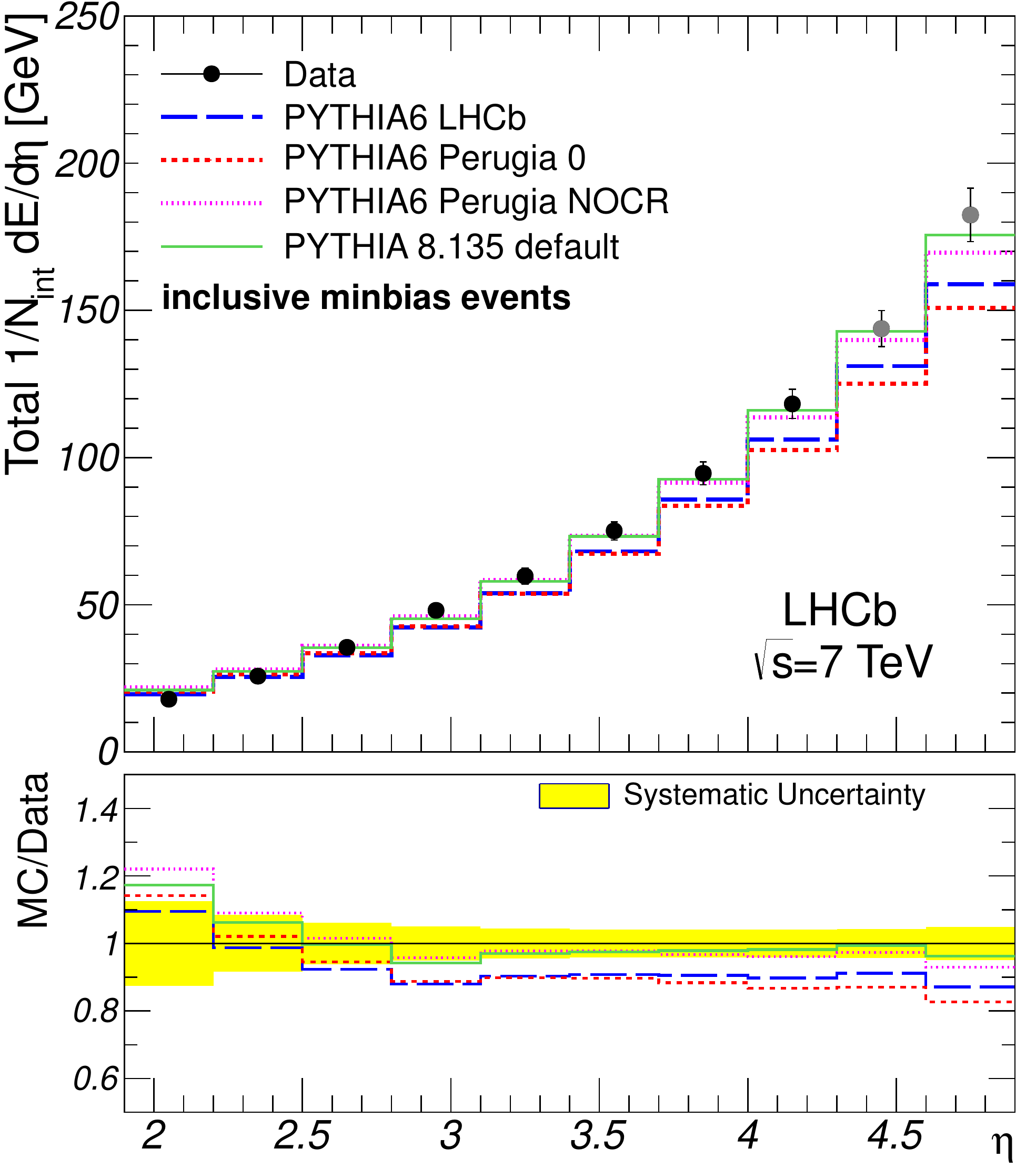}
\includegraphics[width=0.295\textwidth]{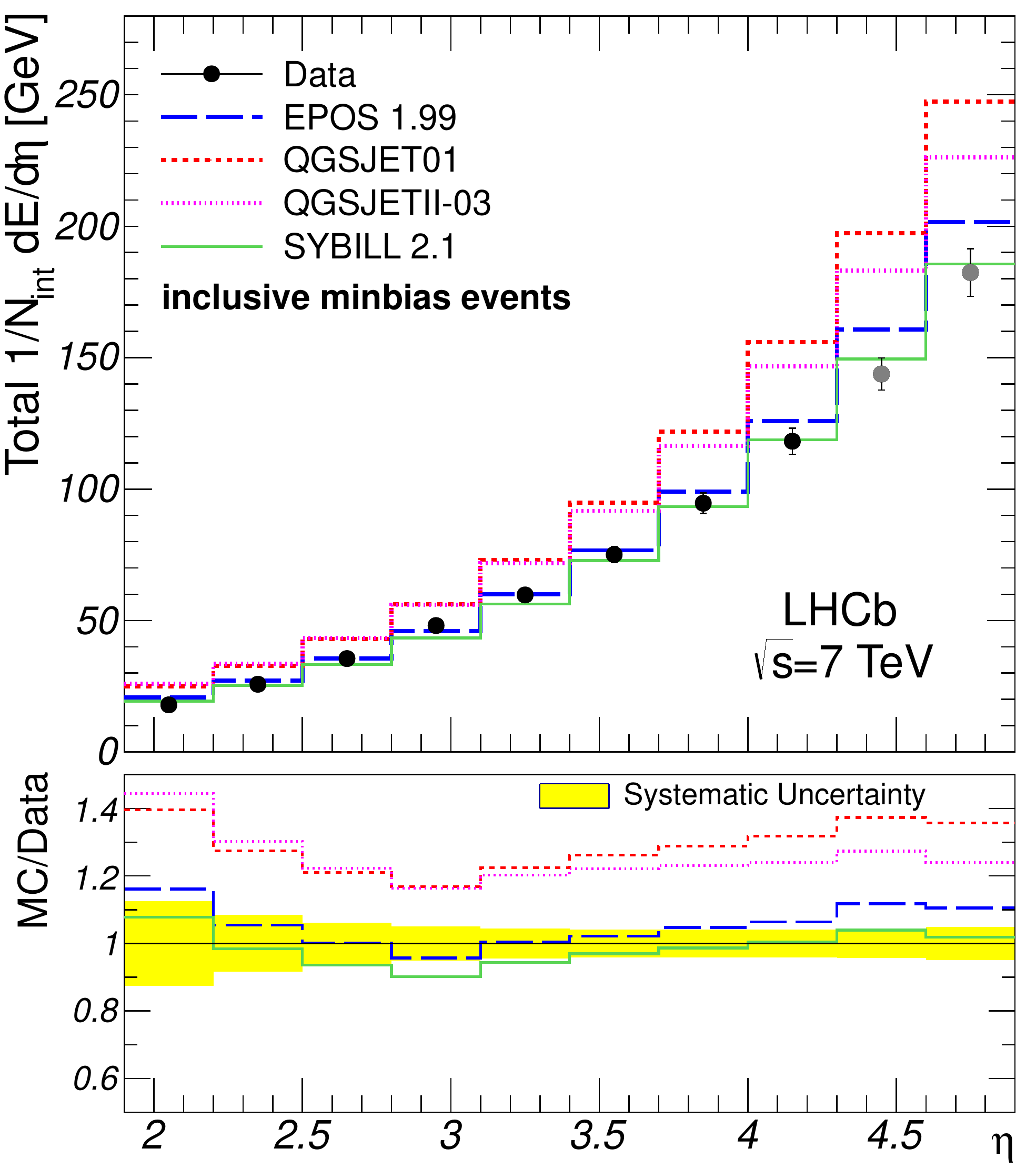}
\caption{\small Total energy flow for inclusive minimum-bias interactions along with predictions 
             given by {\sc Pythia}-based~(left) and cosmic-ray interaction models~(right). 
             The data are indicated by points with error bars representing the systematic uncertainties, 
             while MC predictions are shown as histograms. The statistical uncertainties are negligible. The results for other event classes 
             can be found in Ref.~\cite{EFpaper}.
        }
\label{fig:EF} 
\end{figure}
The latter represent a predominant contribution to the soft component 
of a hadron-hadron collision, called the underlying event.
Its precise theoretical description still remains a challenge. 
To constrain the underlying event models, the energy flow is measured in 
$pp$ collisions at $\sqrt{s}=7$~TeV within the pseudorapidity range 
$1.9<\eta<4.9$ with data recorded by the LHCb detector~\cite{lhcb}.
As described in Ref.~\cite{EFpaper}, the primary measurement is the energy flow carried 
by charged particles, while a data-constrained Monte Carlo~(MC) estimate of the neutral component 
is used for the measurement of the total energy flow.
To probe various aspects of multi-particle 
production in hadron-hadron collisions, 
the measurements are performed for inclusive minimum-bias (containing 
at least one track with $p>2$~GeV/c in $1.9<\eta<4.9$), 
hard scattering (having at least one track with $p_{\rm T}>3$~GeV/c in $1.9<\eta<4.9$), 
diffractive, and non-diffractive enriched interactions. 
The last two event types were selected among the inclusive minimum-bias interactions 
requiring the absence or presence of at least one backward track reconstructed by 
the LHCb Vertex Locator in $-3.5<\eta<-1.5$, respectively. 
Experimental results are compared to predictions given 
by the {\sc Pythia}-based~\cite{Skands:2009zm,Clemencic:LHCbMC,P8} 
and cosmic-ray MC event generators~\cite{Enterria:2011}, 
which model the underlying event activity in different ways.
Though the evolution of the energy flow as a function of $\eta$ 
is reasonably well reproduced by the MC generators, none of the models 
used in this analysis are able to describe the energy flow measurements 
for all event classes that have been studied. 
The majority of the {\sc Pythia} tunes underestimate the measurements at large $\eta$, 
while most of the cosmic-ray interaction models overestimate them as can be seen in Fig.~\ref{fig:EF}.
The energy flow is found to increase with the momentum transfer in an underlying $pp$ inelastic interaction.

\section{Exclusive dimuon production}
\begin{table}[b!]
\begin{tabular}{l|c}
\hline
$\sigma [pp \rightarrow pp~J/\psi(\mu^{+}\mu^{-})]$                       &  $474\pm12\pm51\pm92$~pb    \\
$\sigma [pp \rightarrow pp~\psi(2S)(\mu^{+}\mu^{-})]$                     &  $12.2\pm1.8\pm1.3\pm2.4$~pb  \\
$\sigma [pp \rightarrow pp~\chi_{c0}(\gamma~J/\psi(\mu^{+}\mu^{-}))]$     &  $9.3\pm2.2\pm3.5\pm1.8$~pb     \\
$\sigma [pp \rightarrow pp~\chi_{c1}(\gamma~J/\psi(\mu^{+}\mu^{-}))]$     &  $16.4\pm5.3\pm5.8\pm3.2$~pb     \\
$\sigma [pp \rightarrow pp~\chi_{c2}(\gamma~J/\psi(\mu^{+}\mu^{-}))]$     &  $28.0\pm5.4\pm9.7\pm5.4$~pb     \\
$\sigma [pp \rightarrow pp\mu^{+}\mu^{-}], M_{\mu\mu}>2.5~{\rm GeV}/c^{2}$ &  $67\pm10\pm7\pm15$~pb     \\
  \hline
\end{tabular}
\caption{Cross-section measured in $pp$ collisions at $\sqrt{s}=7$~TeV for different exclusive processes. 
The final state particles are required to have pseudorapity in the range $2.0<\eta<4.5$. The first uncertainty is statistical, 
the second is systematic, and the third is due to the luminosity. See Ref.~\cite{EP} for details. }
\label{tableEP}
\end{table}
In $pp$ collisions exclusive processes are elastic reactions of 
the type $pp \rightarrow ppX$, where the protons remain intact and $X$ 
can either be a resonance or a continuum state created through photon and/or gluon propagators. 
If the latter is involved, exclusive processes allow investigation of pomeron and odderon states. 
These studies can be performed in a clean experimental environment 
when $X$ represents a dimuon final state which can be produced 
via the diphoton process leading to a continuous dimuon invariant mass spectrum, 
or via the photon-pomeron process which can produce $\phi$, $J/\psi$, $\psi(2S)$ 
and $\Upsilon$ family resonances which decay into two muons. 
These processes were studied at LHCb by selecting $pp$ collision events 
with no backwards tracks reconstructed by the Vertex Locator 
and exactly two forward tracks. By adding a photon in the final state, 
exclusive production of $\chi_{c}$ states, occurring via pomeron-pomeron fusion, has also been explored.
First cross-section measurements for exclusive $J/\psi$, $\psi(2S)$, $\chi_{c}$ and 
for non-resonant production $pp \rightarrow pp\mu^{+}\mu^{-}$ are carried out 
with 3~${\rm pb}^{-1}$ of low pile-up data recorded by the LHCb detector at $\sqrt{s}=7$~TeV
and are summarised in Table~\ref{tableEP}. The measured cross-sections are found to be 
in agreement with the corresponding theoretical predictions 
which have large uncertainties~\cite{EP}.

\section{Electroweak boson production}
\begin{figure}[t!]
\centering
\includegraphics[width=0.535\textwidth]{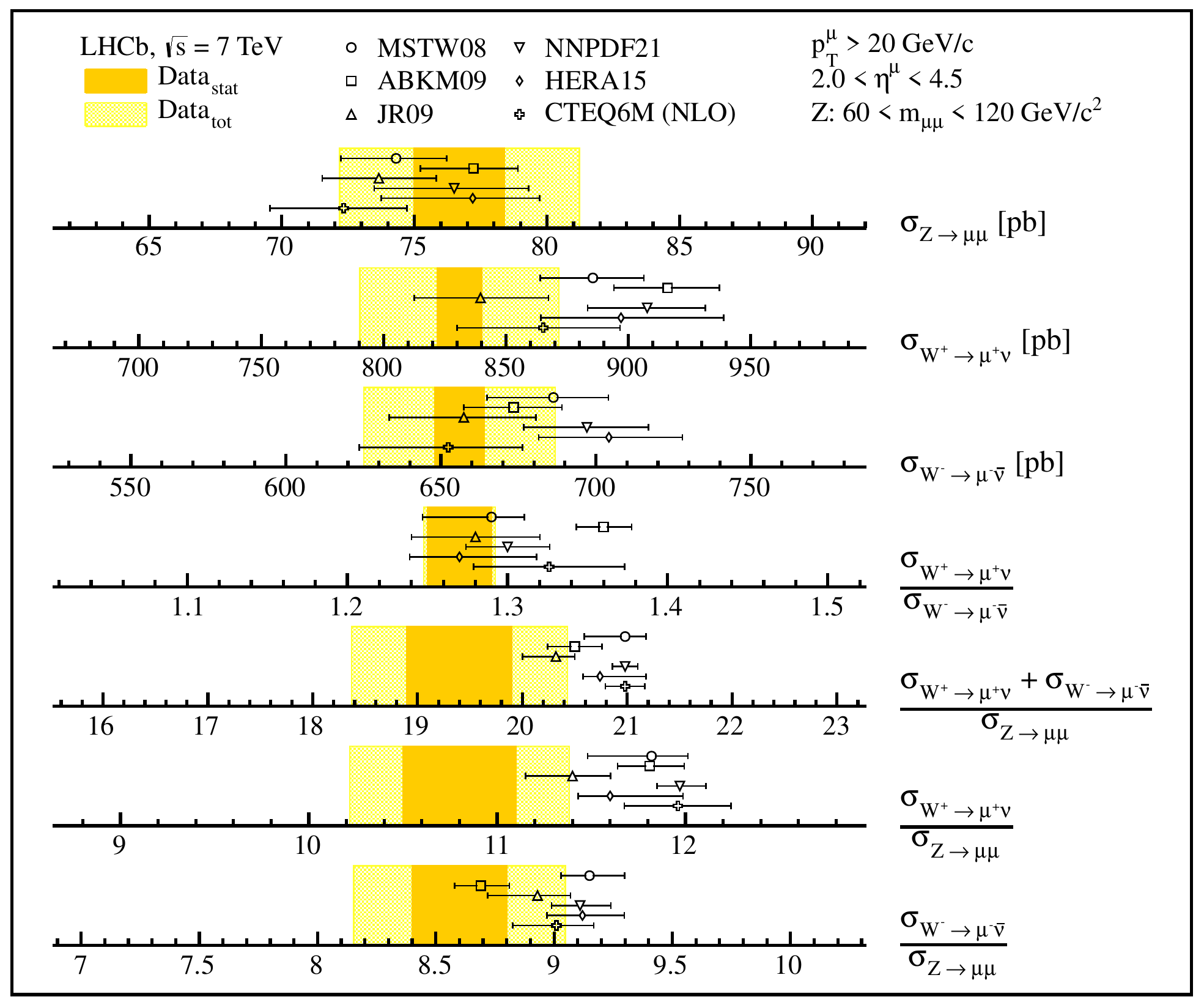}
\includegraphics[width=0.485\textwidth]{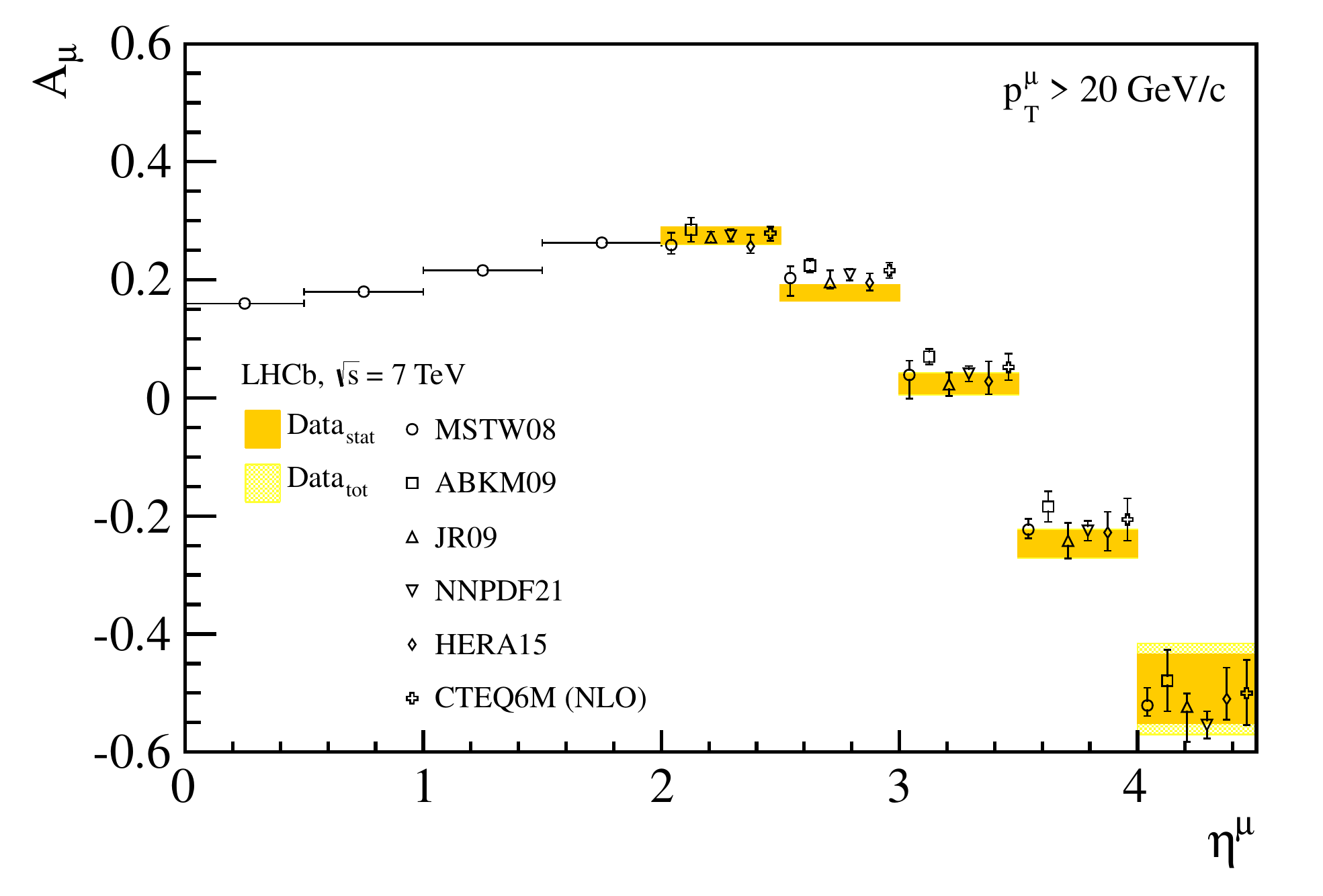}
\caption{\small $Z$, $W^{+}$ and $W^{-}$ cross-section and their ratios~(left) and 
                lepton charge asymmetry $A_{\mu}=(\sigma_{W^{+} \rightarrow \mu^{+}\nu} - \sigma_{W^{-} \rightarrow \mu^{-}\nu}) /
                (\sigma_{W^{+} \rightarrow \mu^{+}\nu} + \sigma_{W^{-} \rightarrow \mu^{-}\nu})$~(right) along with NNLO~(NLO) predictions. 
                The dark shaded (orange) bands indicate the statistical uncertainties, while the light hatched (yellow) bands show 
                the total uncertainties. See Ref.~\cite{EWboson} for details.
        }
\label{fig:EWboson} 
\end{figure}
Measurements of the $W$, $Z$ and low mass Drell-Yan production cross-sections 
in the forward region at LHC energies constitute an important test 
of the Standard Model and provide valuable input to 
the knowledge of the parton density functions of the proton.
These measurements are carried out for the first time within the LHCb acceptance 
using $37~{\rm pb}^{-1}$ of $pp$ collision data recorded at $\sqrt{s}=7$~TeV~\cite{EWboson,DY}. 
The $W$ and $Z$ bosons are reconstructed from muons with $p_{\rm T}>20~{\rm GeV}/c$ and $2.0<\eta<4.5$, 
and, in the case of $Z$, a dimuon invariant mass $M_{\mu\mu}$ between 60 and $120~{\rm GeV}/c^{2}$.
The cross-sections are measured to be $831\pm9\pm27\pm29$~pb for $W^{+}$,
$656\pm8\pm19\pm23$~pb for $W^{-}$, $76.7\pm1.7\pm3.3\pm2.7$~pb for $Z$, 
where the first uncertainty is statistical, the second is systematic 
and the third is due to the luminosity.  The $W$ and $Z$ cross-section ratios and 
the lepton charge asymmetry are also measured in the same kinematic region. 
Figure~\ref{fig:EWboson}~(left) shows the cross-sections and their ratios 
along with NNLO QCD predictions with 6 different sets for the parton density functions. 
The lepton charge asymmetry, which has a strong $\eta$ dependence in the forward region, 
is shown in Fig.~\ref{fig:EWboson}~(right). All $W$ and $Z$ measurements are found to be 
in general agreement with the NNLO predictions. 
Further cross-section measurements are performed 
with $Z \to \tau\tau$ and $Z \to e^{+}e^{-}$ channels~\cite{Ztautau,Zee}. 
\begin{figure}[t!]
\centering
\includegraphics[width=0.65\textwidth]{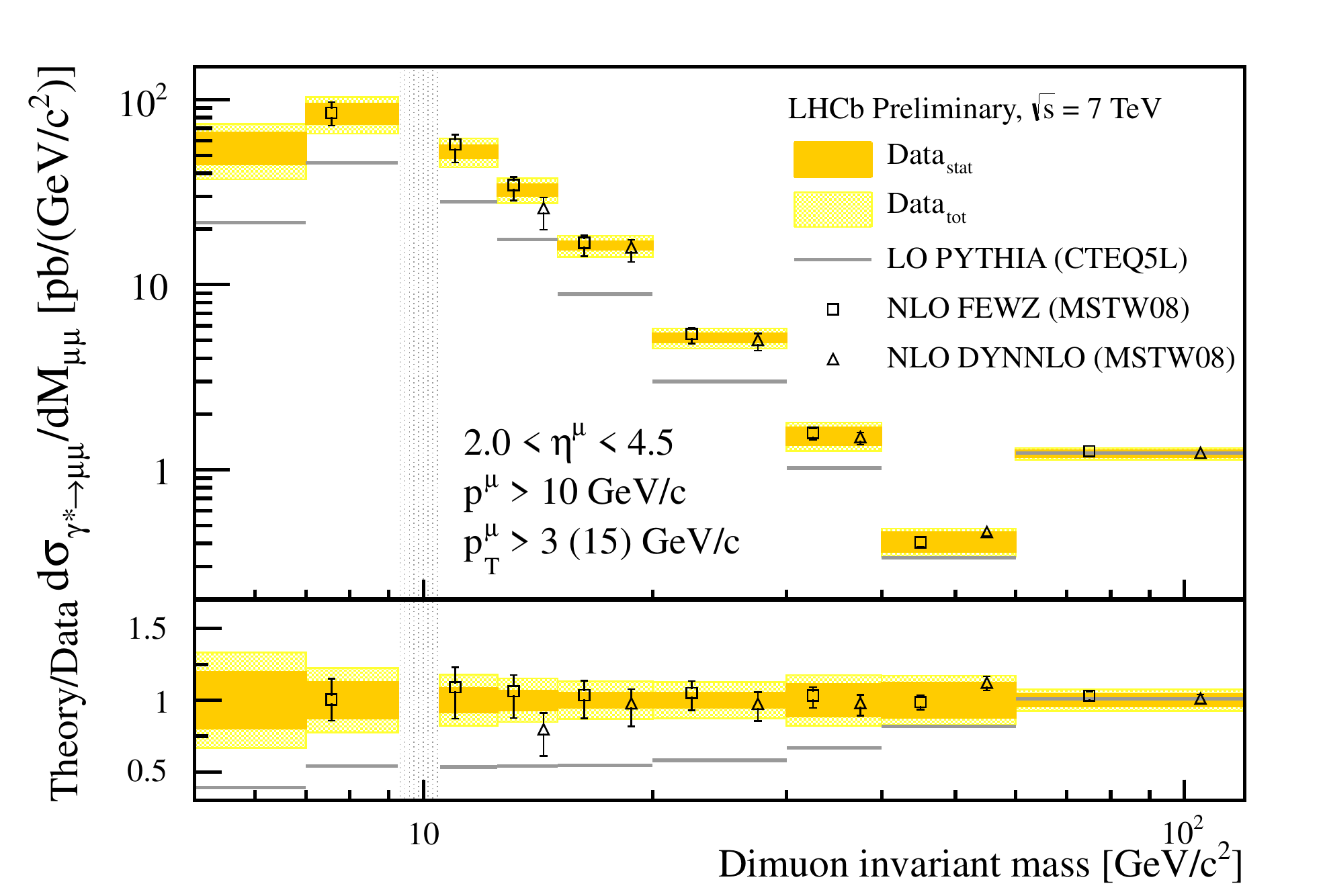}
\caption{\small Differential cross-section for $\gamma^{*} \rightarrow \mu\mu$ as a function of $M_{\mu\mu}$.
         The dark shaded (orange) bands indicate the statistical uncertainties, while the light shaded (yellow) 
         bands show the total uncertainties. The shaded vertical band corresponds to the excluded mass region of the $\Upsilon$ meson. 
         Superimposed are the {\sc Pythia} and NLO predictions as described in Ref.~\cite{DY}.
        }
\label{fig:DY} 
\end{figure}

Low mass Drell-Yan dimuon production is studied in the mass range 
$5<M_{\mu\mu}<120~{\rm GeV}/c^{2}$. Figure~\ref{fig:DY} illustrates its differential cross-section 
as a function of $M_{\mu\mu}$ for muons with $2.0<\eta<4.5$, $p>10~{\rm GeV}/c$ and $p_{\rm T}>3~{\rm GeV}/c$ 
($p_{\rm T}>15~{\rm GeV}/c$ for $M_{\mu\mu}>40~{\rm GeV}/c^{2}$). 
The data are found to be in reasonable agreement with two different NLO predictions, 
while the LO {\sc Pythia} underestimates the measurements for low $M_{\mu\mu}$.

\section{Summary}
In addition to a rich heavy flavour physics programme,
the LHCb experiment performs important studies 
of QCD and electroweak processes in a unique kinematic range.
These measurements provide a sensitive test of the Standard Model 
delivering valuable input for theoretical models.
         

%

{\raggedright
\begin{footnotesize}



\bibliographystyle{aipproc}   




\end{footnotesize}
}

\end{document}